# A small shoulder of optical absorption in polycrystalline $HfO_2$ by LDA+U approach

Jinping Li,[1,4,a)] Songhe Meng,[1] Liyuan Qin,[1] Hantao Lu,[2,4] and Takami Tohyama[3,4]

[1] Center for Composite Materials and Structures, Harbin Institute of Technology, Harbin 150080, China

[2] Center for Interdisciplinary Studies & Key Laboratory for Magnetism and Magnetic Materials of the MoE, Lanzhou University, Lanzhou 730000, China

[3] Department of Applied Physics, Tokyo University of Science, Tokyo 125-8585, Japan

[4] Yukawa Institute for Theoretical Physics, Kyoto University, Kyoto 606-8502, Japan

The dielectric function of the wide-gap optical material $HfO_2$ is investigated by the local density approximation (LDA) $+U$ approach. We focus on the origin of the shoulder-like structure near the edge of the band gap in the imaginary part of the dielectric function, which has been observed on the thin films of monoclinic $HfO_2$. A comparison study on the three polymorphs of hafnia shows that regardless of the underlying crystal structure, the existence of the shoulder is directly controlled by the value of the shortest length of Hf-O bonds. The proposition is further supported by the numerical simulations of isostatic pressing. A possible implication in high-pressure measurements is suggested accordingly.

## I. INTRODUCTION

Hafnium dioxide ($HfO_2$) is widely used in optical and protective coatings, capacitors, and phase shifting masks etc., as one of the most promising high dielectric constant materials.[1] It is well known that the hafnia has three polymorphs at atmospheric pressure[2]: the monoclinic, the tetragonal and the cubic, denoted as m-, t-and c-$HfO_2$, respectively. Regardless of numerous studies on its optical properties[3,4], some issues remain open. For the thin films of m-$HfO_2$ grown on amorphous silica substrates, a small shoulder at the threshold of the absorption spectra is observed,[5] which has aroused considerable interest. While in other two phases no similar

---

a) Electronic mail: lijinping@hit.edu.cn

structure has been reported. From the first-principles calculation it has been suggested that the shoulder structure might be an intrinsic property of crystallized monoclinic $HfO_2$, as a consequence of the reconstruction of the electronic states near the band edge following the adjustment of the crystal structure.[6] In this paper, we re-examine the issue in wider perspective by investigating the electronic structures and optical properties of $HfO_2$ across these three phases with adjustable cell parameters. More specifically, in a given phase, we have modified (amplify or decrease) the cell parameters homogeneously with respect to their experimental values without changing the underlying lattice structure. The LDA+U results show that different from the previous proposition,[6] the emergence of the shoulder-like structure in the dielectric function is actually determined by the shortest Hf-O bond length (denoted by Ds). When t- and c-$HfO_2$ are squeezed to Ds=0.20±0.01 nm (7% changes), the shoulder structures appear; on the other hand, in the phase of m-$HfO_2$, the increase of the cell parameters up to 7% with Ds=0.22 nm destroys the shoulder. We reach the conclusion that as a key factor, it is Ds, not the crystal structure, that determines the optical absorption near the band edge. Further simulations in high-pressure environments support the argument.

## II. COMPUTATIONAL METHODOLOGY

The density functional theory calculations are performed with plane-wave ultrasoft pseudopotential, by using the LDA with CA-PZ functional and the LDA+U approach as implemented in the CASTEP code (Cambridge Sequential Total Energy Package).[7] The ionic cores are represented by ultrasoft pseudopotential for Hf and O atoms. For Hf atom, the configuration is [Xe] $4f^{14}5d^{2}6s^{2}$, where the $5d^2$ and $6s^2$ electrons are explicitly treated as valence electrons. For O atom, the configuration is [He] $2s^{2}2p^{4}$, where $2s^2$ and $2p^4$ electrons are explicitly treated as valence electrons. The plane-wave cut off energy is 380eV and the Brillouin-zone integration is performed over the 24×24×24 grid sizes using the Monkorst-Pack method. This set of parameters assure the total energy convergence of $5.0×10-6$ eV/atom, the maximum force of 0.01eV/Å, the maximum stress of 0.02GPa and the maximum displacement of $5.0×10^{-4}$Å. In the LDA+U approach, we set Ud=8.0, 8.25, 8.25 eV and Up=6.0, 6.3, 6.25 eV for m-, t- and c-$HfO_2$, respectively.[6] The details of the method, i.e., the LDA+Ud+Up, can be found in Refs. 6, 8 It is noted that for the situations with prescribed cell parameters, energy calculations without structural optimization are applied. While for the calculations in variant isostatic pressing, the electronic structure and optical properties are obtained after structural optimization.

## III. RESULTS AND DISCUSSION

### A. The main results of three polymorphs of $HfO_2$ with varying cell parameters

The numerical results are summarized in Table I. The experimental values of Ds in m-, t- and c-$HfO_2$ are 0.2039, 0.2066 and 0.2200nm, respectively (No. 1-3 entries in Table I).[9] The Ds obtained by LDA +U after structural optimization are about 7% difference from the experiments on average, [6] as shown in No. 4-6 entries in Table I. The results without structural optimization with prescribed cell parameters are shown in No. 7-9 entries. In No. 7 (m1), Ds for m-$HfO_2$ is 7.878% larger than the experimental value (from 0.2039 to 0.2200nm), and the shoulder disappears (see Fig.1 (a)). While in No. 8 (t1) and No. 9 (c1), Ds for t- and c-$HfO_2$ are set to 7.33% smaller (from 0.2066 to 0.1915nm, and from 0.2200 to 0.2038nm, respectively), where the shoulder structure appears (Fig.1 (a) again). Further, the critical value of Ds in the numerical simulation seems to coincide to the same value (around 0.21 to 0.22nm), regardless of the underlying crystal structures. The last three entries, i.e., No. 10-12, are numerical results with structural optimization in variant isostatic pressures. In details, No. 10 (m2) shows the results of m-$HfO_2$ under -100GPa pressure, with the change ratio of 5.145% with respect to the value in No. 4 (both are obtained with structural optimization being applied). In this stretching case, the shoulder vanishes. On the other hand, No. 11 (t2) and 12 (c2) are about t- and c-$HfO_2$ under 100GPa pressures, respectively. In these pressing cases, the shoulders are observed (see Fig. 5 in the text). As will be shown below.

TABLE I. The relations between cell parameters, Ds and shoulder existence in polycrystalline $HfO_2$ in various settings. No.1-3, experimental values; No.4-6, numerical results after structural optimization; No.7-9, results obtained by energy calculations without structural optimization; No.10, results for m-$HfO_2$ under -100GPa with structural optimization; No.11 and 12, results for t- and c-$HfO_2$ under 100GPa with structural optimization. Note that the change ratios in No. 4-9 are calculated with respect to the corresponding (experimental) values in No. 1-3; while the change ratios in No. 10-12 are set to be the variations with respect to those in No. 4-6, since they are all numerical results with structural optimization being applied.

| No. | Phases | Band gap (eV) | Cell parameters (nm) | Ds (nm) | Change ratio (%) | Shoulder or not? |
|---|---|---|---|---|---|---|
| 1 | mexp | | 0.5117/0.5175/0.5291/99.2 ° | 0.2039 | - | |
| 2 | texp | | 0.364/0.529 | 0.2066 | - | |
| 3 | cexp | | 0.508 | 0.2200 | - | |
| 4 | m | 5.70 | 0.5386/0.5331/0.5492/99.6 ° | 0.2138 | 4.855% | Yes |
| 5 | t | 5.70 | 0.3767/0.5328 | 0.2307 | 11.665% | No |
| 6 | c | 5.697 | 0.526 | 0.2311 | 5.045% | No |
| 7 | m1 | 5.363 | 0.5520/0.5582/0.5714/99.2 ° | 0.2200 | 7.878% | No |
| 8 | t1 | 6.169 | 0.3373/0.4902 | 0.1915 | -7.335% | Yes |
| 9 | c1 | 5.054 | 0.4707 | 0.2038 | -7.333% | Yes |
| 10 | m2 | 4.719 | 0.5633/0.5701/0.5809/99.5 ° | 0.2248 | 5.145% | No |
| 11 | t2 | 5.161 | 0.3429/0.4850 | 0.2100 | -8.973% | Yes |
| 12 | c2 | 5.164 | 0.4893 | 0.2100 | -9.130% | Yes |

Figure 1 shows the imaginary part of the dielectric functions and total density of states (DOS) for three polymorphs of $HfO_2$ with prescribed cell parameters obtained by LDA+Ud+Up (corresponding to t1-, c1-, and m1-$HfO_2$ in Table I). From Fig. 1(a), we can see that small shoulders emerge in the imaginary part of the dielectric function, if Ds in t- and c-$HfO_2$ are set to be smaller. Note that the shoulder in t1 is pushed to higher energy (over than 6.5 eV), which is largely due to the fact that the gap value in t-HfO2 is greatly enhanced (from 5.70 eV to 6.169 eV) by the decrease of Ds. On the other hand, with Ds in m-$HfO_2$ enlarged to 0.22nm, the shoulder structure disappears. The total DOS are presented in Fig. 1(b) as a reference. From Fig. 1(b), we can notice relatively complicated structures of DOS of valence bands for t1 and c1 near the Fermi energy.

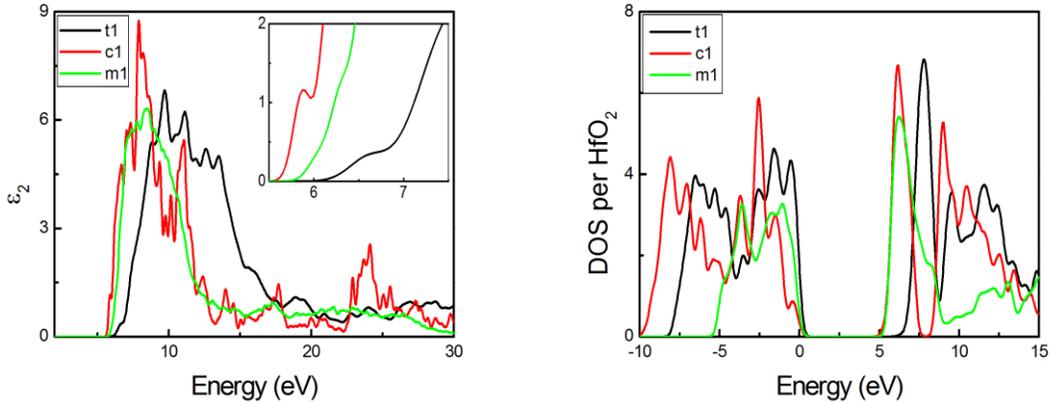

FIG. 1 (a) The imaginary part of the dielectric functions and (b) total DOS for t1-, c1- and m1-HfO$_2$ with changing Ds obtained by LDA+Ud+Up. The inset in (a) displays the details of ε2 near the gap edge.

## B. The DOS and band structures of three polymorphs of HfO$_2$ with varying cell parameters

In order to better understand the effects of varying Ds on the electronic structure and the subsequent optical properties, we need to compare the total DOS of HfO$_2$ with prescribed cell parameters to the ones obtained by structural optimization. Figure 2 shows the total DOS of m1- (with larger Ds) and m-HfO$_2$, and of t1- (with less Ds) and t-HfO$_2$, obtained by LDA+Ud+Up. The DOS difference between c1- and c-HfO$_2$ is similar to the tetragonal case and not shown here. It is clear from Fig. 2 that by compressing the crystal homogeneously (decreasing Ds in the numerical simulations, corresponding to m1-to-m and t-to-t1 in Fig. 2), the band structures near the band-gap edges become more dispersed in general, together with some variations in the band gap (more information on the band gap values can be read in Table I).

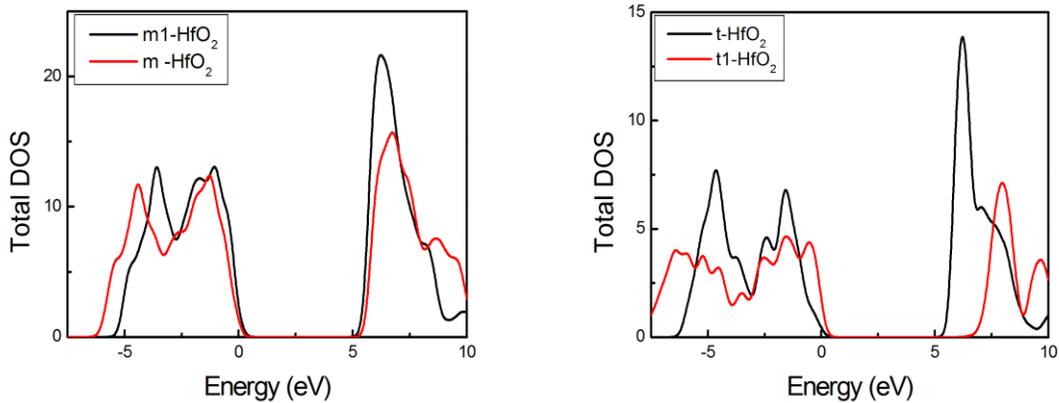

FIG. 2 The total DOS of m1- and m-HfO$_2$ (a), t1- and t-HfO$_2$ (b), obtained by LDA+Ud+Up.

The tendency of band-width broadening around the Fermi surface by decreasing Ds can also be demonstrated by the detailed band structure calculations. In Figs. 3, 4, the band structures around the Fermi level for m- and m1-, t- and t1- HfO$_2$ obtained by LDA+Ud+Up are shown, where due to the Ds shortening (m1-to-m and t-to-t1), the broadening of the band width near the Fermi energy and the less concentrated available states in the conduction bands are clear recognized. Subsequently, on top of the main absorption peak, it produces more diverse structures into the optical spectrum, e.g., the small shoulder around 6 eV. Reversely, stretching Ds induces the narrowing of the band width around the Fermi energy, and the resulting denser available states in the conduction bands can wipe out the shoulder structure in the optical absorption, as demonstrated in the m-to-m1 case. Moreover, the evolution of the band structure is largely determined by Ds, i.e., the shortest Hf-O bond length. Numerically, there seems to be a critical value for Ds (around 0.21 to 0.22nm). Smaller than that, the shoulder-structure can emerge, regardless of the underlying crystal phases.

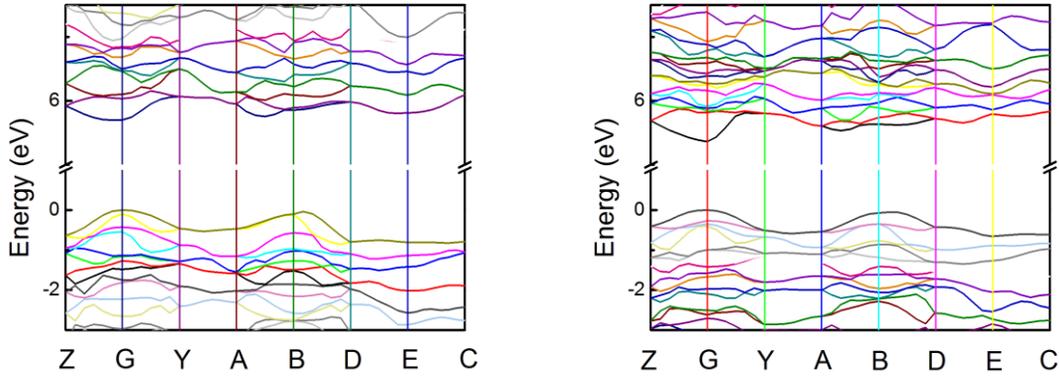

FIG. 3 The band structures of (a) m-$HfO_2$ and (b) m1-$HfO_2$ near the Fermi surface obtained by LDA+Ud+Up.

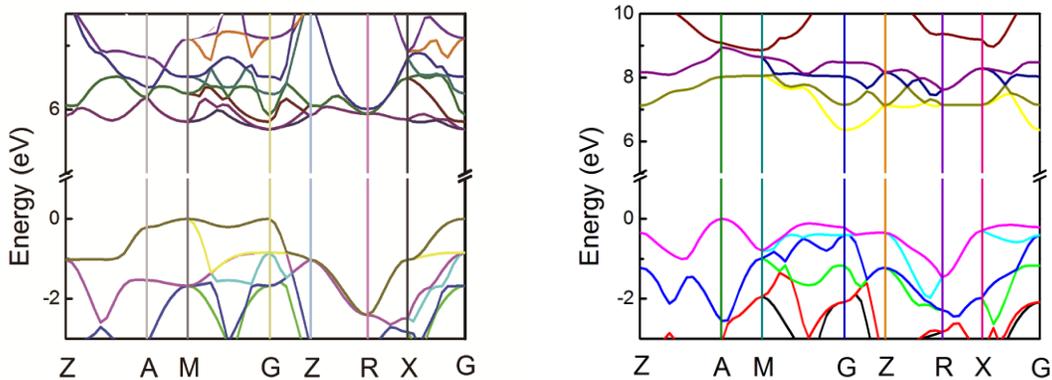

FIG. 4 The band structures of (a) t-$HfO_2$ and (b) t1-$HfO_2$ near the Fermi surface obtained by LDA+Ud+Up.

## C. The tetragonal and cubic $HfO_2$ under high pressures

Based on the above speculations, we suggest that with appropriate pressure applied, the shoulder-like structure in the imaginary part of the dielectric function, similar to that in the normal state of monoclinic $HfO_2$, can also be observed in tetragonal and cubic phases. We hence perform further numerical simulations for c-HfO$_2$ and t-HfO$_2$ under cold isostatic pressing. In Fig.5, the results obtained by LDA+Ud+Up show that small shoulder-like structures indeed appear in the imaginary part of the dielectric functions in t2-HfO$_2$ and c2-HfO$_2$ under 100GPa isostatic pressing. Based on these observations, we predict the existence of a shoulder-like structure in the absorption spectra near the band edge for both t- and c-HfO$_2$ subjected to high isostatic pressures.

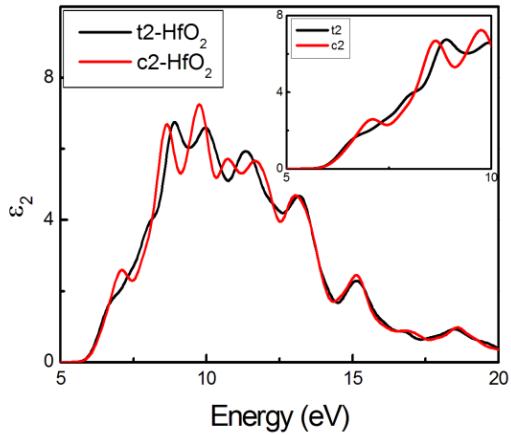

FIG.5 The imaginary part of the dielectric functions in t2-HfO$_2$ and c2-HfO$_2$ under 100GPa isostatic pressing obtained by LDA+Ud +Up approach. The inset displays the details of ε2 near the band edge.

## IV. CONCLUSIONS

In summary, the dielectric function of HfO$_2$ in its three polymorphs has been revisited by LDA+Ud+Up approach. We find that keeping the lattice structure intact, the electronic band structure and the consequent optical properties are heavily influenced by small changes of Ds, the shortest Hf-O bond length. When Ds is lower than a critical value, the shoulder-like structure near the edge of the band gap in the imaginary part of the dielectric function emerges, regardless the underlying crystal structures. This observation might provide new insights into the optical properties of this wide-gap material. Further theoretical results obtained by LDA+U approach show that small shoulders do appear in the absorption spectra when the tetragonal and the cubic HfO$_2$ are subjected to 100GPa isostatic pressing. A possible implication in high-pressure measurements is proposed.

Acknowledgements

This work was supported by the Foundation for Innovative Research Groups of the National Natural Science Foundation of China (Grant No. 11121061, and 11474136). The authors thank Prof. Xiaoguang Luo in China Academy of Aerospace Aerodynamics, for his valuable discussions.